\documentclass[pre,a4paper,superscriptaddress,nofootinbib]{revtex4}

\usepackage{graphicx}
\usepackage[body={16cm,22cm},bottom=1.5cm]{geometry}
\usepackage{amsmath}
\usepackage{latexsym}
\usepackage{amsfonts}
\usepackage{psfrag}

\begin{document}
\title{Modelling transient absorption and thermal conductivity in a
simple nanofluid}
\author{Mihail Vladkov \footnote{E-mail: mihail.vladkov@lpmcn.univ-lyon1.fr},
Jean-Louis Barrat \footnote{E-mail:
jean-louis.barrat@lpmcn.univ-lyon1.fr}}

\affiliation{
  Laboratoire de Physique de la Matière Condensée et Nanostructures
               Université Lyon 1; CNRS; UMR 5586
               Domaine Scientifique de la Doua
               F-69622 Villeurbanne cedex; France}

\date{\today}
\setcounter{page}{1}

\begin{abstract}
Molecular dynamics simulations are used to simulate the thermal
properties of a model fluid containing nanoparticles (nanofluid).
By modelling transient absorption experiments, we show that they
provide a reliable determination of interfacial resistance between
the particle and the fluid. The flexibility of molecular
simulation allows us to consider separately the effect of
confinement, particle mass and Brownian motion on the thermal
transfer between fluid and particle. Finally, we show that in the
absence of collective effects, the heat conductivity of the
nanofluid is well described by the classical Maxwell Garnet
equation model.
\end{abstract}

\maketitle

\newpage
Many experimental studies have suggested that the thermal
conductivity of colloidal suspensions referred to as
``nanofluids'' is unusually high \cite{Eastman,Patel}.
Predictions of effective medium theories are accurate in some cases \cite{Putnam-poly} but
generally fail to account for the large enhancement in conductivity.
In spite of a large number of - sometimes conflicting or
controversial - suggestions and experimental findings \cite{Keblinski-Cahill},  the
microscopic mechanisms for such an increase remain unclear.
Among the possibilities that were suggested, the Brownian
motion \cite{prasher} of a single sphere in a liquid leads to an
increase in thermal conductivity of the order of $4-5\%$, and
appears to be an attractive and generic explanation. The essential
idea is that the Brownian velocity of the suspended particle
induces a fluctuating hydrodynamic flow
\cite{Alder,Keblinski-flow}, which on average influences
(increases)
 thermal transport. This mechanism is different from transport of
 heat through center of mass diffusion, which was previously shown
 to be negligible \cite{Keblinski}. However, some recent
experimental high precision studies reported a normal conductivity
in nanoparticle suspensions at very small volume fractions below
1\% \cite{putnam}, questioning the validity of this assumption.

In  this work we use nonequilibrium molecular dynamics
"experiments" to explore further the transfer of heat in a model
fluid containing nanoparticles. Our approach is closely related to
experimental techniques, but we also make use of the flexibility
allowed by molecular simulations to explore extreme cases in terms
e.g. particle/fluid mass density mismatch. We concentrate on
model systems that are expected to be representative of generic
properties.

We start our study by mimicking the "pump-probe" experiments that
are used in nanofluids to estimate interfacial resistance, which
is an essential ingredient in modelling the thermal properties of
highly dispersed system \cite{pump-probe-Cahill}. All atoms in our
system interact through Lennard-Jones interactions
\begin{equation}
\label{eq:ljpotcut} U_{lj}(r) =  \bigg\{ \begin{array}{lll}
         4\varepsilon((\sigma/r)^{12} - c(\sigma/r)^{6}), &r\le r_c \\
         0, &r>r_c
       \end{array}
\end{equation}
where $r_c=2.5\sigma$. The coefficient $c$ is equal to 1 for atoms
belonging to the same phase, but can be adjusted to modify the
wetting properties of the liquid on the solid particle
\cite{barratbocquet,barratchiaruttini}. Within the solid particle,
atoms are linked with their neighbors through a
  FENE
(Finite extension non-linear elastic) bonding potential:
\begin{equation}
\label{eq:FENE} U_{FENE}(r) =  \frac{k}{2}R_0
\ln(1-(\frac{r}{R_0})^2), \qquad r<R_0
\end{equation}
where $R_0=1.5\sigma$ and $k=30.0 \varepsilon/\sigma^{2}$. The
solid particle in the fluid was prepared as follows: starting from
a FCC bulk arrangement of atoms at zero temperature, the atoms
within a sphere were linked to their first neighbors by the FENE
bond. Then the system was equilibrated in a constant NVE ensemble
with energy value corresponding to a temperature $T=1$. The
particle contains 555 atoms, surrounded by 30000 atoms of liquid.
The number density  in  the system is $\rho = 0.85\sigma^{-3}$.
Taking $\sigma=0.3nm$ this corresponds to a particle radius of
order $R_{part} \sim 1.5nm$ and to a system size $L \sim 10nm$.

The transient absorption simulation starts with an equilibrium
configuration at temperature $T=1$, by "heating" uniformly the
nanoparticle.  This heating is achieved by rescaling the
velocities of all atoms within the solid particles, so that the
kinetic energy per atom is equal to $3\epsilon$. We then monitor
the kinetic energy per atom of the
particle as a function of time, which we take
as a measure of the particle temperature.  The system evolves at
constant energy, but the average temperature of the liquid, which
acts essentially as a reservoir, is only very weekly affected by the
cooling process. Within a few time steps the kinetic temperature
of the  particle drops to a value of $T_p \approx 2$. This
evolution corresponds to the standard one for an isolated,
harmonic system. As the particle was equilibrated at $T_p = 1$, we
have due to kinetic and potential energy equipartition $\langle
E_{pot}(t=0) \rangle = 1/2$. As we start our simulation with
$\langle E_k(t=0) \rangle = 3$, within a very short time the
kinetic energy drops to a value of $1.5$, then the potential
energy stored in the particle atoms positions yields its
contribution of $1/2$ to the temperature, equilibrating it to a
value of $2$. This first step does not involve any heat exchange
with the liquid surroundings.

\begin{figure}[ht]
\centering
\includegraphics[width=7cm]{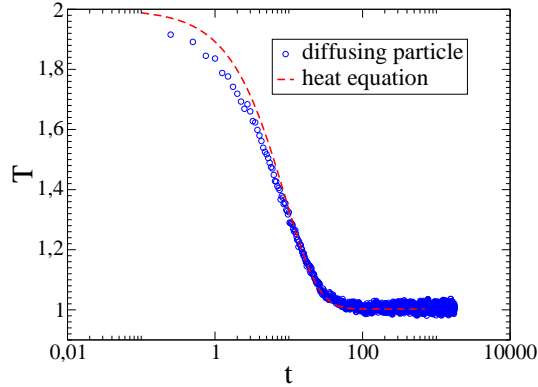}
\caption[Comparison of simulation results and calculation]
{Comparison between the temperature evolution from simulations and
the solution of the continuum heat equation. The value of the
Kapitza resistance taken for the calculation is $R_K = 0.8$.}
\label{fig:fit}
\end{figure}

 The subsequent decrease of the particle temperature, on
  the other hand, directly
probes such exchanges. A quantitative understanding of this decay
is particularly important, as it remains an essential experimental
tool to quantify heat transfer across the particle-liquid
interface. In figure \ref{fig:fit}, we compare the molecular
dynamics simulation result for the temperature as a function of
time, to the result of a continuum calculation involving the
interfacial (Kapitza) thermal resistance as an adjustable
parameter. The continuum calculation makes use of the standard
heat transfer equations
\begin{eqnarray}
C\frac{dT_p}{dt} &=& -4\pi R_{p}^{2} j(R_p, t) \\
\frac{\partial T_l}{\partial t} &=&
D_{th}\frac{1}{r^2}\frac{\partial}{\partial r}\bigg(
r^2\frac{\partial T_l(r,t)}{\partial r}\bigg)
\end{eqnarray}
where $T_l(r,t)$ and $T_p$ are the liquid and particle
temperatures, respectively.  $C$ is the thermal capacity of the
particle, $R_p$ its radius, $D_{th}$ is the thermal diffusion
coefficient of the liquid. The above equations are solved with the
following boundary conditions:
\begin{eqnarray}
j(R_p,t) &=& \frac{1}{R_{k}}\big(T_l(R_p^+,t) - T_p(t)\big) \\
j(R_{\infty},t) &=& 0 \label{eq:boundary}
\end{eqnarray}
where $R_{\infty}$ is chosen so that $\frac{4}{3} \pi
R_{\infty}^3$ is equal to the volume of the simulation box from
the previous section. The initial condition is
\begin{eqnarray}
T_p(0) &=& 2 \\
T_l(r,0) &=& 1
\end{eqnarray}
The temperature was assumed to be uniform inside the nanoparticle.
This assumption is based on the simulation results, where the
observed temperature profile inside the particle was found
independent of position within statistical accuracy. We used data
found in the literature \cite{barratchiaruttini, palmer} for the
values of the fluid thermal diffusivity and conductivity. As the
simulated particle is not exactly spherical, but  presents some FCC
facets, its radius for use in the calculation was estimated from
the radius of gyration:
\begin{equation}
\langle R_g^2 \rangle = \frac{1}{N} \sum_1^N (r_i - r_{CM})^2 =
\frac{3}{5}R_p^2
\end{equation}
where $R_g^2$ is the measured radius of gyration of the particle
atoms, and the second equality applies to an ideal sphere. In the
range $T=1$ to $T=3.5$, we checked through equilibrium simulations
that the   heat capacity of the particle is very close to $3k_B T
N$, as for an harmonic ideal solid.

The value of the interface thermal resistance (Kapitza resistance)
appearing in equation \ref{eq:boundary} was adjusted to fit the simulation
data. The value that fits the simulation results for the wetting
system $m=1$ and $c=1$ was found to be $R_K \approx 0.8$. This
number is in agreement with the thermal resistance for a wetting
flat wall calculated in \cite{barratchiaruttini} for a similar
system with a different potential in the solid phase, and
a completely different simulation method.

The same cooling simulation was performed using a non wetting
particle ($c=0.5$). A substantial   slowing down of the cooling
rate was also observed, which can  be attributed to an increased
Kapitza resistance. The resulting value of $R_K$ is 3.2
(Lennard-Jones unit), again in
 agreement with previous determinations for flat surfaces
\cite{barratchiaruttini}. In real units, a value $R_K=1$
corresponds typically to an interfacial \textit{conductance}
$G=1/R_K$, of the order of $100\mathrm{ MW/K m}^2$
\footnote{The conversion to physical units is made by taking
a Lennard-Jones time unit $\tau_{LJ}=10^{-12}s$, and a
length unit $\sigma=0.3nm$. The unit for $G$ is
$energy/temperature/(length)^2/time$. As the
energy/temperature ratio is given by the Boltzmann
constant $k_B$, we end up with a unit for $G$ equal
to $k_B /\sigma^2 /\tau_{LJ} \simeq =10^8 W/m^2/s$}.
The method is therefore a sensitive probe of interfacial resistance,
as usually assumed in experiments.

\begin{figure}[ht]
\centering
\includegraphics[width=7cm]{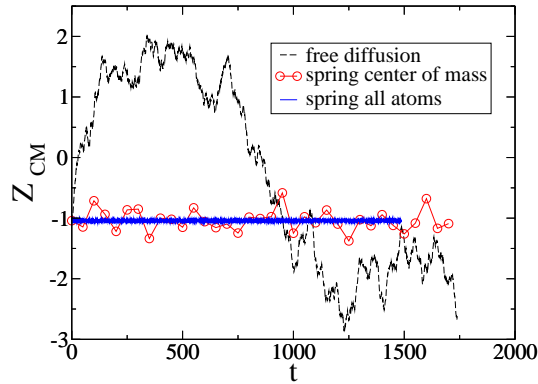}
\caption[Particle center of mass evolution] {Evolution of the $Z$
coordinate of the particle center of mass for the different
systems: free particle, particle confined by a spring attached to
its center of mass and particle where all atoms are tethered to
their initial position. The spring constant for all springs is
$k=30$.} \label{fig:com}
\end{figure}

In a second step, we explore the influence of thermal Brownian
motion of the particle on the cooling process. First, let us
recall that the naive idea, that diffusion could speed up cooling
by displacing the particles towards cooler fluid regions is easily
excluded. Quantitatively, diffusion of the particle and heat
diffusion take place on very different time scales. The diffusion
coefficient of the particle was measured, in our case, to be three
orders of magnitude smaller than the  heat diffusion coefficient
in the fluid. We also suppressed diffusion by tethering the
particle to its initial position with a harmonic spring of
stiffness $k=30\epsilon/\sigma^2$. As expected, no effect is
observable on the cooling rate. This measurement cannot probe for
another possible effect - the influence of fluid flow on the
cooling. As discussed in \cite{Acrivos} the heat transfer from  a
sphere in a low Reynolds number velocity field is enhanced by the
latter. Because of the diffusion velocity of the Brownian particle
$v = \big( \frac{3k_B T}{m}\big)^{1/2}$, it can be viewed, at any
given moment, as a particle in a velocity field \cite{prasher}. To
probe the influence of this phenomenon, we tether every single
atom in the particle to its initial position with a harmonic
spring ($k=30$) and compare the measured temperature evolution
with the previous results. The cooling rate is still not
influenced by this manipulation even if the center of mass is now
``frozen'' (fig. \ref{fig:com}).

\begin{figure}[ht]
\centering
\includegraphics[width=7cm]{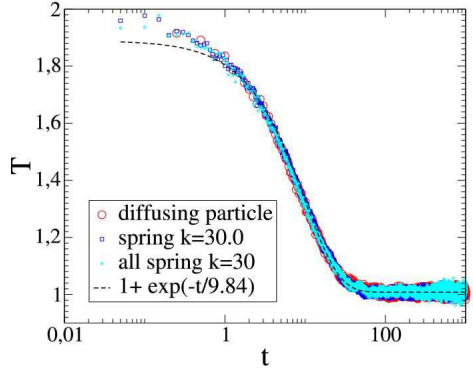}
\caption[Effect of confinement on cooling]
{The evolution of the particle temperature
for the different systems: free particle, particle confined by a spring
attached to its center of mass and particle where all atoms are tethered to
their initial position. Every curve is the mean value from 20 simulation runs.
No effect is observed.}
\label{fig:spring}
\end{figure}

 A final check on the influence
 of such velocity effects was attempted  by modifying the mass
of the atoms that constitute the nanoparticle. This artificial
procedure reduces the thermal Brownian velocity, and when it is
carried out we indeed observe a strong slowing down of the cooling
process. However, this slowing down is again completely
independent of the center of mass motion of the particle, which is
controlled by the presence of the tethering springs. On the other
hand, the effect of this mass density increase is easily
understood in terms of an increase of the interfacial resistance.
 A higher  mass of the particle atoms
decreases the speed of sound in the solid and thus leads to a
larger  acoustic mismatch between the two media, which slows down
the cooling. Numerically, we find that for a mass of $100$ times the
mass of a liquid atom, the
Kapitza resistance increases to $R_K=7.4$.

In summary, we have shown that the Brownian motion of the particle
does not affect the cooling process. As a byproduct, we have shown
that the mass density parameter provides a flexible numerical way
of tuning the interfacial resistance, which will be used in the
next section.
\begin{figure}[ht]
\centering
\includegraphics[width=7cm]{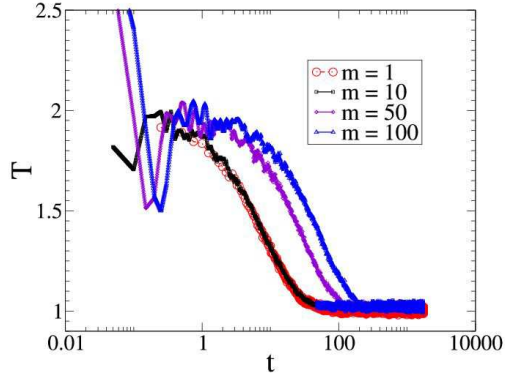}
\caption[Effect of particle mass on cooling] {The evolution of the
particle temperature as a function of the mass of the particle
atoms. Every curve is the mean value from 20 simulation runs. The
increased mass slows down heat exchange.} \label{fig:mass}
\end{figure}

\begin{figure}[ht]
\centering
\includegraphics[width=7cm]{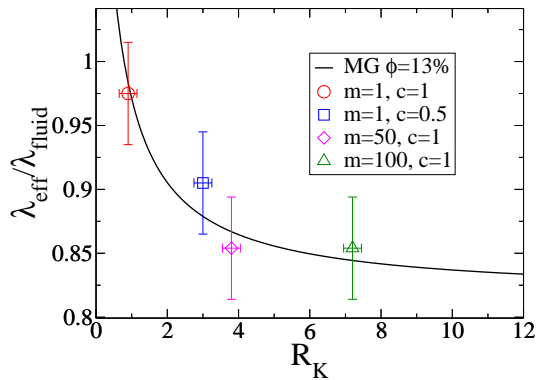}
\caption[Comparison of simulation results and MG equation]
{Comparison between the ratio of the effective conductivity to the
conductivity of the pure liquid of the simulated systems and the
values obtained from the Maxwell Garnett equation. The values of
the Kapitza resistance for the simulations were obtained from the
cooling rate of the particle.} \label{fig:mg}
\end{figure}

In this last section, we attempt direct measurements of the
nanofluid heat conductivity using a nonequilibrium molecular
dynamics simulation of heat transfer across a fluid slab
containing one nanoparticle, with periodic boundary conditions in
the $x$ and $y$ direction and confined by a flat repulsive potential in
the $z$ direction.
\begin{figure}[ht]
\centering
\includegraphics[width=6cm]{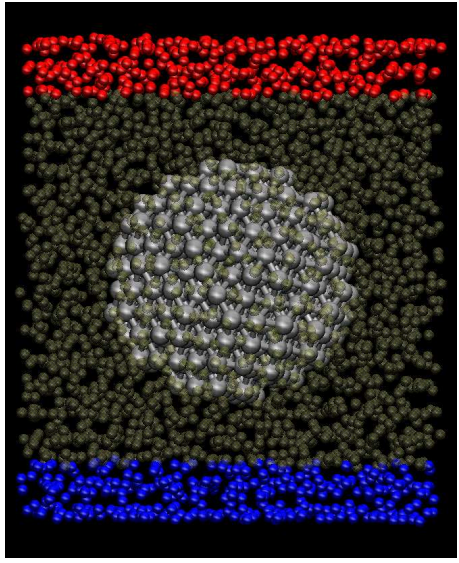}
\caption[Snapshot of heat transfer setup] {Snapshot of the
system used to evaluate the thermal conductivity with a
particle of 13\% volume fraction.} \label{fig:snap}
\end{figure}
Two slices of fluid are thermostated, using velocity rescaling,
at different temperatures. To avoid any effect of thermophoresis
or coupling of the thermostat the particle, the particle is
constrained to stay at equal distance between the two thermostats
using two different schemes. First, a confinement between two
repulsive, parallel  walls,  that couple only to the particle
atoms. The particle is then constrained to the mid-plane
of the simulation cell, but free to diffuse within this plane. The
second possibility is to tether the center of mass to a fixed
point, as described above, so that any possible effect due to flow
or diffusion is eliminated. The energy flux is measured by
calculating the energy absorbed by the thermostats. The effective
conductivity of the system was defined as:
\begin{equation}
\lambda_{eff} = \frac{J}{(T_1 - T_2)/L}
\end{equation}
The temperatures of the two thermostats were $T_1 = 2$ and $T_2 =
1$. In order to compare the conductivity results for the different
systems they were first equilibrated to the same pressure at a
temperature $T=1.5$, then a non equilibrium run was performed for
about 1500-2000 $\tau_{LJ}$ to make sure the pressure stays the
same for the systems of different nature and finally a production
run of about 30000 $\tau_{LJ}$ during which thermostats energy,
particle diffusion and temperature profiles are monitored.
Simulations were performed with two different volume fractions for
the particle, 2\% or 13\%. The volume fraction is defined as the
volume of the particle divided by the volume of the fluid
outside the thermostats.
  As expected from the study above, no effect of the particle
diffusion on the fluid conductivity was observed. The effective
conductivity measured for the particle diffusing in 2D, the
particle attached with a single spring or the particle where all
atoms were attached to their initial positions has the same value
within 1\% which is below the error bar of the measurement (around
4-5\%). Finally,  we investigated the effect of the presence of
the nanoparticle on the thermal conductivity of the fluid. For the
smallest volume fraction ($\Phi \sim 2\%$), we were not able to
detect any change in thermal conductivity compared to the bulk
fluid.  At the higher volume fraction ($\Phi \sim 13\%$), on the
other hand, we observe a clear {\it decrease} in the heat
conductivity associated with the presence of the nanoparticle
(fig. \ref{fig:mg}). Clearly, this decrease must be interpreted in terms
of interfacial effects. To quantify these effects, we use the
 Maxwell-Garnett approximation for
spherical particles, modified to  account account for the Kapitza
resistance at the boundary between the two media. The resulting
expression for the effective conductivity \cite{MG} is
\begin{equation}
\frac{\lambda_{eff}}{\lambda_l}= \frac{\big(
\frac{\lambda_p}{\lambda_l}(1+2\alpha)+2 \big) + 2\Phi\big(
\frac{\lambda_p}{\lambda_l}(1-\alpha)-1 \big)} {\big(
\frac{\lambda_p}{\lambda_l}(1+2\alpha)+2 \big) - \Phi\big(
\frac{\lambda_p}{\lambda_l}(1-\alpha)-1 \big)}
\end{equation}
where $\lambda_l$ and $\lambda_p$ are the liquid and particle
conductivities, $\Phi$ is the particle volume fraction and $\alpha
= \frac{R_K \lambda_l}{R_p}$ is the ratio between the Kapitza
length (equivalent thermal thickness of the interface) and the
particle radius. This model predicts an increase in the effective
conductivity  for $\alpha
> 1$  and a
decrease for $\alpha < 1$, regardless of the value of the
conductivity of the particles or of the  volume fraction. The
prediction depends very weekly on the ratio $\lambda_p/\lambda_l$,
less than 1\% for $10<\lambda_p/\lambda_l<100$.
 The minimum value   of
$\lambda_{eff}/\lambda_l$, obtained when $\alpha \to \infty$, is
$\frac{1-\Phi}{1+\Phi/2}$ while the    maximum possible
enhancement (for $\lambda_p \to \infty$ and $R_K \to 0$) is
$\frac{1+2\Phi}{1-\Phi}$.

 As explained above, the Kapitza
resistance can be modified  by tuning either the liquid solid
interaction coefficient $c$, or the mass density of the solid, or
a combination thereof. Figure \ref{fig:mg} illustrates the
variation of the measured effective conductivity for several
values of the Kapitza resistance, determined independently for
various values of these parameters. It is seen that the observed
variation (decrease in our case) in the effective conductivity is
very well described by the Maxwell-Garnett expression.
This expression also allows us to understand why the heat
conductivity does not vary in a perceptible manner for the smaller
volume fraction, for which the predicted change would be less than
$3\%$, within our statistical accuracy.

\section*{Conclusion}
We have explored some important aspects of the thermal properties
of "nanofluids", at the level of model system and individual solid
particles. The molecular modeling of transient heating experiments
confirms that they are a sensitive tool for the determination of
thermal boundary resistances. The effect of Brownian motion on the
cooling process, on the other hand, was found to be  negligible.
By varying interaction parameters or mass density, we are able to
vary the interfacial resistance between the particle and the fluid
in a large range. This allowed us to estimate, over a large range
of parameters, the effective heat conductivity of a model
nanofluid in which the particles would be perfectly dispersed. Our
results can be simply explained in terms of the classical
Maxwell-Garnett model, provided  the interfacial resistance is
taken into account. The essential parameter that influences the
effective conductivity turns out to be the ratio between the
Kapitza length and the particle radius, and for very small
particles a decrease in conductivity compared to bulk fluids is
found. We conclude that large heat transfer enhancements observed in
nanofluids must originate from collective effects, possibly
involving particle clustering and percolation or cooperative heat
transfer modes.

\bibliographystyle{unsrt}

\end{document}